# Statistical adjustment for a measure of healthy lifestyle doesn't yield the truth about hormone therapy

Diana B. Petitti[*,1] and Wansu Chen[*,2]

*University of Southern California and Kaiser Permanente Southern California*

**Abstract:** The Women's Health Initiative randomized clinical trial of hormone therapy found no benefit of hormones in preventive cardiovascular disease, a finding in striking contrast with a large body of observational research. Understanding whether better methodology and/or statistical adjustment might have prevented the erroneous conclusions of observational research is important. This is a re-analysis of data from a case-control study examining the relationship of postmenopausal hormone therapy and the risks of myocardial infarction (MI) and ischemic stroke in which we reported no overall increase or decrease in the risk of either event. Variables measuring health behavior/lifestyle that are not likely to be causally with the risks of MI and stroke (e.g., sunscreen use) were included in multivariate analysis along with traditional confounders (age, hypertension, diabetes, smoking, body mass index, ethnicity, education, prior coronary heart disease for MI and prior stroke/TIA for stroke) to determine whether adjustment for the health behavior/lifestyle variables could reproduce or bring the results closer to the findings in a large and definitive randomized clinical trial of hormone therapy, the Women's Health Initiative.

For both MI and stroke, measures of health behavior/lifestyle were associated with odds ratios (ORs) less than 1.0. Adjustment for traditional cardiovascular disease confounders did not alter the magnitude of the ORs for MI or stroke. Addition of a subset of these variables selected using stepwise regression to the final MI or stroke models along with the traditional cardiovascular disease confounders moved the ORs for estrogen and estrogen/progestin use closer to values observed in the Women Health Initiative clinical trial, but did not reliably reproduce the clinical trial results for these two endpoints.

## 1. Background

The Women's Health Initiative (WHI) clinical trial of hormone therapy is a large randomized trial whose primary aim was determining whether post-menopausal hormone therapy prevents coronary heart disease (Rossouw et al. [10] and Anderson et al. [1]). The study began in the early 1990's and published main results in 2002–2004. It involved recruitment and randomization of more than 18,000 postmenopausal women to hormones or placebo. The WHI found no overall effect, or

[*]Supported by grant number R01-HL-47043 from the National Heart Lung and Blood Institute.
[1]University of Southern California, Department of Preventive Medicine, Keck School of Medicine, 299 E. Laurel Avenue, Sierra Madre, CA 91024, USA, e-mail: dbpetitti@verizon.net
[2]Kaiser Permanente Southern California, Department of Research and Evaluation, 100 S. Los Robles, Pasadena, CA 91101, USA, e-mail: Wansu.Chen@kp.org

*AMS 2000 subject classifications:* Primary 92C60; secondary 62P10, 00B30.

*Keywords and phrases:* cerebrovascular disorders, cerebral infarction, stroke, coronary, epidemiological methods, estrogen, heart disease, hormone replacement, myocardial infarction.

perhaps an increase, in the risk of myocardial infarction (MI) in women assigned to combined estrogen/progestin (E/P) therapy and no effect of estrogen alone (E) (Rossouw et al. [10] and Anderson et al. [1]). Prior observational research concluded that the risk of coronary heart disease was reduced by half (Stampfer and Colditz [12]). Moreover, the risk of stroke was increased for both E and E/P in WHI (Rossouw et al. [10] and Anderson et al. [1]). Prior observational research found no effect of hormone therapy on stroke (Psaty et al. [9]).

Publications by Prentice et al. suggest statistical approaches that would have overcome the discrepancy between the observational research and the clinical trial (Prentice et al. [7] and Prentice et al. [8]). Reviews of these approaches are mixed (Petitti and Freedman [5], Freedman and Petitti [3], DeMets [2] and Greenland [4]). Understanding whether better methodology and/or statistical adjustment might have prevented the erroneous conclusions made based on the observational research is important.

We previously published the results of a case-control study in which we estimated the relative risks of myocardial infarction (MI) and stroke in current users of E and E/P that adjusted for traditional cardiovascular disease confounders—age, ethnicity, education as a measure of socioeconomic status, and factors known to increase the risk of MI or stroke causally (smoking, diabetes, hypertension, body mass index), using logistic regression (Petitti et al. [6] and Sidney et al. [11]). The study neither confirmed nor ruled out a lower risk of MI or stroke in hormone users. Adjusted estimates of the relative risk of MI and stroke were both, however, less than 1.0 in current E and EP users compared with never hormone users, a finding inconsistent with WHI.

This is a re-analysis of data from our case-control study. In the re-analysis, variables that assess health behaviors/lifestyle were added to the multivariate analysis along with the traditional confounders to determine whether adjustment for these variables, as a marker for healthy lifestyle, could reproduce or bring the results closer to the findings in the Women's Health Initiative for stroke and MI. Some of the health behavior/lifestyle variables were chosen specifically because they were NOT likely to be causally related to MI or stroke and whose relationship with these conditions would be expected to be non-causal.

## 2. Methods

### 2.1. Overview

Study methods and results examining the association of hormone therapy with the risk of stroke and MI after adjustment for traditional confounders are described in detail in two prior publications (Petitti et al. [6] and Sidney et al. [11]). Briefly, an attempt was made to identify all fatal and non-fatal strokes and MIs in women age 45-74 years in 10 medical centers of the Kaiser Permanente Medical Care Program, Northern California region, during the period, November, 1991 – November, 1994. A standard protocol was used to define stroke and subclassify it by type. Our re-analysis includes only strokes classified as ischemic.

For each case, an attempt was made to interview one control, matched on year of birth and facility of usual care. Out of 550 ischemic stroke cases and 685 MI cases, we were able to obtain 349 stroke case/controls sets and 438 MI case/control sets for analysis, after exclusions described in the Analysis section.

*2.2. Information*

Eligible cases and controls were interviewed in-person by trained interviewers using a standardized instrument. Interview questions were asked relative to an index date, which was the date of symptom onset for cases and the same date for her matched control. If a case had died or was unable to communicate verbally, an attempt was made to interview a proxy, but proxy responses are excluded from this analysis.

Hypertension was defined as a "yes" to a question about use of medication for high blood pressure. Diabetes was defined as a "yes" to a question asking about use of insulin or pills for diabetes. Women who stated that they had been told by a physician that they had a high cholesterol level were classified as having hypercholesterolemia (high cholesterol level was not used as a covariate in the original analysis). A study subject was defined as a nonsmoker if she answered "no" to the question, "Have you ever smoked cigarettes?" If she answered "yes" to this question, she was categorized as a current regular smoker on the basis of her answer to the question, "On [index date] were you still smoking regularly?" (Regularly means at least 5 cigarettes per week, almost every week). Body mass index was calculated from self-reported height and weight.

The questions asked about health behavior/lifestyle spanned a range of activities and behaviors that people believe may improve their health. We sought especially to identify questions about activities and behaviors that were NOT related causally to cardiovascular disease risk. Women were asked, "Do you do the following to try to improve your health?" This was followed by the trained interviewer reading each behavior with a query for a "yes" or "no" response.

*2.3. Analysis*

We defined current hormone use as "yes" for hysterectomized women who used E within 1 month of the index date; current hormone use was defined as "yes" for non-hysterectomized women who used E/P within 1 month of the index date; and "no" otherwise. Excluded from the analysis were pre-menopausal women and hysterectomized women who used EP, non-hysterectomized women who used E, and users of progestin only within 1 month of the index. All exclusions were applied in the same way to cases and controls. These exclusions and definitions are the same as in our prior published analyses.

The odds ratio (OR) was used to estimate the relative risk of MI and stroke. The multivariate analysis used conditional logistic regression analysis. Ninety-five percent confidence intervals (CI) were calculated for OR estimates.

We first calculated age-adjusted ORs for MI and stroke in relation to the traditional vascular disease risk factors and in relation to use of E and E/P separately in hysterectomized and non-hysterectomized women. We then examined ORs for stroke and MI in relation to each of the lifestyle/behavior questions adjusting first for age only and then for age and the traditional cardiovascular disease confounders, also separately in women with and without a hysterectomy. Last, we used stepwise logistic regression to screen the list of behavior/lifestyle variables and select only those meeting a significance level of 0.2 or lower for both entering and staying in the models while forcing all the traditional vascular disease risk factors to stay in the model. Thus, only those behavior/lifestyle variables deemed important statistically to the outcomes are included (separate analyses for women with and without a hysterectomy).

### 3. Results

Although we attempted to reproduce the study samples using the same exclusion criteria described in the two publications, we were unable to gather six stroke case/control sets and four MI case/control sets. Tables 1 and 2 show the characteristics of non-hysterectomized and hysterectomized cases and controls and age-adjusted ORs and 95% C.I.s for MI and stroke in relation to the traditional confounders and current use of E/P (for non-hysterectomized women) and E only (for hysterectomized women). The traditional cardiovascular disease risk factors show expected associations with the risk of MI and stroke.

Tables 3 and 4 show ORs and 95% C.I.s for stroke and MI in relation to each of the variables that measure health behavior in non-hysterectomized and hysterectomized women adjusting for age and then for age and the traditional confounders. For almost all of the health behavior questions, age-adjusted ORs for both stroke and MI in women who responded "yes" to the question are lower than 1.0 in both non-hysterectomized and hysterectomized women. Adjustment for the traditional cardiovascular disease risk factors in addition to age does not change the magnitude of the OR for any variable by much, although the C.I.s often include 1.0.

The fully adjusted ORs for MI in non-hysterectomized women who stated they regularly used sunblock or sunscreen (adjusted OR 0.3; 95% C.I. 0.2–0.5) or who stated they were trying to cut down on alcohol use (adjusted OR 0.4; 95% C.I. 0.2–0.7) are of particular note. The fully adjusted OR for stroke in hysterectomized women who stated they regularly used sunblock or sunscreen (OR 0.4; 95% C.I. 0.2–0.9) is also noteworthy.

The ORs in women who answered yes to the behavior questions are further from 1.0 for MI than for stroke in both non-hysterectomized and hysterectomized women (Tables 3–4). The adjustment for the traditional confounders changes the estimates less for MI than for stroke in both non-hysterectomized and hysterectomized women.

Table 5 shows the results of models assessing the ORs for stroke and MI in relation to current use of E and E/P after adjustment. The ORs for stroke reported in the previous publication were based on the sample that includes women with hysterectomy who used E/P and women without hysterectomy who used E and therefore are not exactly the same as the results in this re-analysis. The ORs were estimated for current use of E and E/P after adjusting for age only, for age and the traditional confounders, and then for age and the traditional confounders plus the behavior/lifestyle variables that were selected using the stepwise regression procedure. Estimates for the risk of coronary heart disease and stroke from WHI in non-hysterectomized users of E/P and hysterecomized users of E are shown for comparison. Adjustment for the behavior/lifestyle variables in addition to the traditional confounders results in further changes in the ORs for current use of E and E/P. However, neither adjustment for the traditional confounders nor adjustment for the behavior/lifestyle variables in addition to the traditional confounders reliably reproduces the WHI results considering both disease endpoints and both E and E/P.

### 4. Discussion

For both MI and stroke, the variables that measured healthy behavior/lifestyle were associated with ORs less than 1.0 even after adjustment for age. Even after adjustment for the traditional cardiovascular disease confounders in addition to age, the ORs for the variables that measured healthy behavior/lifestyle remained mostly

TABLE 1
*Women without hysterectomy: for demographic and other characteristics, percent of cases and controls and age adjusted odds ratios for myocardial infarction and stroke for each characteristic*

| Characteristic | Myocardial infarction | | | Stroke | | |
|---|---|---|---|---|---|---|
| | Cases (N = 189) | Controls (N = 199) | OR (95% CI) | Cases (N = 156) | Controls (N = 153) | OR (95% CI) |
| Current E/P, % | 20.1 | 29.7 | 0.6 (0.4–1.01)* | 18 | 24.2 | 0.7 (0.4–1.2)* |
| Treated for hypertension, % | 41.3 | 30 | 1.6 (1.1–2.5) | 49.7 | 27.5 | 2.6 (1.6–4.2) |
| Treated for diabetes, % | 22.3 | 8.1 | 3.3 (1.8–6.1) | 27.1 | 11.8 | 2.8 (1.5–5.1) |
| Body Mass Index | | | | | | |
| Quartile 1 (lowest) | 24.9 | 30.3 | 1.0 (ref) | 23.8 | 30.5 | 1.0 (ref) |
| Quartile 2 | 17.8 | 28.8 | 0.7 (0.4–1.3) | 22.5 | 24.5 | 1.2 (0.6–2.3) |
| Quartile 3 | 25.4 | 17.2 | 1.8 (0.98–3.2) | 29.2 | 23.2 | 1.6 (0.9–3.1) |
| Quartile 4 | 31.9 | 23.7 | 1.6 (0.96–2.8) | 24.5 | 21.8 | 1.4 (0.8–2.7) |
| Cigarette smoking % | | | | | | |
| Never | 45 | 50.2 | 1.0 (ref) | 35.3 | 53.6 | 1.0 (ref) |
| Past | 21.7 | 33.7 | 0.7 (0.5–1.2) | 30.1 | 32 | 1.5 (0.9–2.5) |
| Occasional/Current | 33.3 | 16.1 | 2.5 (1.5–4.3) | 34.6 | 14.4 | 3.8 (2.1–6.9) |
| Race/Ethnicity % | | | | | | |
| White, non-Hispanic | 76.8 | 75.7 | 1.0 (ref) | 66 | 77.8 | 1.0 (ref) |
| Hispanic | 8.9 | 7.1 | 1.2 (0.6–2.6) | 11.1 | 8.5 | 1.6 (0.7–3.4) |
| African-American | 5.4 | 7.6 | 0.7 (0.3–1.7) | 12.4 | 4.6 | 3.3 (1.3–8.2) |
| Asian | 5.4 | 7.6 | 0.7 (0.3–1.7) | 8.5 | 6.5 | 1.6 (0.7–3.9) |
| Other or unknown | 3.8 | 2 | 1.9 (0.5–6.6) | 2 | 2.6 | 0.9 (0.2–4.0) |
| Level of Education % | | | | | | |
| Less than high school | 17.9 | 11.7 | 1.0 (ref) | 22.2 | 12.4 | 1.0 (ref) |
| High school graduate | 35.9 | 33 | 0.7 (0.4–1.3) | 33.3 | 31.4 | 0.6 (0.3–1.2) |
| Some college or business or technical training | 33.2 | 35.5 | 0.6 (0.3–1.1) | 37.3 | 34 | 0.6 (0.3–1.2) |
| College graduate | 13 | 19.8 | 0.4 (0.2–0.9) | 7.2 | 22.2 | 0.2 (0.1–0.4) |
| History of CHD % | 5.3 | 3.5 | 1.5 (0.5–4.0) | – | – | |
| Prior Stroke / TIA % | – | – | | 5.1 | 0.7 | 8.3 (1.03–67.0) |

* Reference group "Never"

TABLE 2

*Women with hysterectomy: for demographic and other characteristics, percent of cases and controls and age adjusted odds ratios for myocardial infarction and stroke for each characteristic*

| Characteristic | Myocardial infarction | | | Stroke | | |
|---|---|---|---|---|---|---|
| | Cases (N = 125) | Controls (N = 122) | OR (95% CI) | Cases (N = 90) | Controls (N = 85) | OR (95% CI) |
| Current E, % | 68 | 74.6 | 0.7 (0.4–1.2)* | 68.9 | 74.1 | 0.7 (0.4–1.4) |
| Treated for hypertension, % | 47.2 | 34.4 | 1.7 (1.04–2.9) | 46.7 | 27.1 | 2.6 (1.4–5.0) |
| Treated for diabetes, % | 25.6 | 9.8 | 3.2 (1.5–6.5) | 28.1 | 3.5 | 10.5 (3.0–36.3) |
| Body Mass Index | | | | | | |
| Quartile 1 (lowest) | 22.6 | 24.8 | 1.0 (ref) | 20.6 | 28.3 | 1.0 (ref) |
| Quartile 2 | 25 | 32.2 | 0.9 (0.4–1.7) | 30.7 | 23.5 | 1.8 (0.8–4.1) |
| Quartile 3 | 29 | 23.2 | 1.4 (0.7–2.8) | 23.9 | 18.8 | 1.8 (0.7–4.4) |
| Quartile 4 | 23.4 | 19.8 | 1.3 (0.6–2.7) | 25 | 29.4 | 1.2 (0.5–2.7) |
| Cigarette smoking % | | | | | | |
| Never | 37.6 | 50 | 1.0 (ref) | 41.1 | 47 | 1.0 (ref) |
| Past | 30.4 | 33.6 | 1.2 (0.7–2.2) | 31.1 | 41.2 | 0.9 (0.4–1.7) |
| Occasional/Current | 32 | 16.4 | 2.8 (1.4–5.5) | 27.8 | 11.8 | 2.6 (1.1–6.2) |
| Race/Ethnicity % | | | | | | |
| White, non-Hispanic | 87.9 | 82 | 1.0 (ref) | 73.9 | 84.7 | 1.0 (ref) |
| Hispanic | 4.9 | 8.2 | 0.6 (0.2–1.6) | 11.4 | 4.7 | 2.7 (0.8–9.1) |
| African-American | 3.2 | 4.1 | 0.7 (0.2–2.8) | 5.7 | 3.5 | 1.8 (0.4–8.0) |
| Asian | 3.2 | 4.9 | 0.6 (0.2–2.2) | 3.4 | 3.5 | 1.1 (0.2–5.5) |
| Other or unknown | 0.8 | 0.8 | 0.9 (0.1–14.6) | 5.7 | 3.5 | 1.8 (0.4–7.8) |
| Level of Education % | | | | | | |
| Less than high school | 21 | 14.8 | 1.0 (ref) | 26.1 | 5.9 | 1.0 (ref) |
| High school graduate | 37.9 | 29.5 | 0.9 (0.4–1.9) | 30.7 | 28.2 | 0.2 (0.1–0.7) |
| Some college or business or technical training | 29.8 | 33.6 | 0.6 (0.3–1.3) | 31.8 | 43.5 | 0.2 (0.1–0.5) |
| College graduate | 11.3 | 22.1 | 0.4 (0.1–0.9) | 11.4 | 22.4 | 0.1 (0.03–0.4) |
| History of CHD % | 6.4 | 4.9 | 1.3 (0.4–3.9) | – | – | |
| Prior Stroke / TIA % | – | – | | 1.1 | 1.2 | 1.1 (0.1 – 19.0) |

*Reference group "Never".

TABLE 3
*Women without hysterectomy: percent of cases and controls who reported the behavior and adjusted odds ratios for myocardial infarction and stroke in women who reported the behavior*

| Health behaviors | Myocardial infarction | | | | Stroke | | | |
|---|---|---|---|---|---|---|---|---|
| | Cases ($N = 189$) | Controls ($N = 199$) | Age adjusted OR (95% CI) | Adjusted for traditional confounders[1] OR (95% CI) | Cases ($N = 156$) | Controls ($N = 153$) | Age adjusted OR (95% CI) | Adjusted for traditional confounders[2] OR (95% CI) |
| Exercise | 54.6 | 69.2 | 0.5 (0.3–0.8) | 0.6 (0.4–0.99) | 53.6 | 69.9 | 0.5 (0.3–0.8) | 0.6 (0.3–1.0) |
| Try to eat more foods containing fiber | 75 | 85.4 | 0.5 (0.3–0.9) | 0.6 (0.3–0.97) | 73.7 | 86.3 | 0.4 (0.2–0.8) | 0.6 (0.3–1.1) |
| Try to eat foods low in fat | 68.1 | 84.3 | 0.4 (0.2–0.7) | 0.4 (0.2–0.7) | 76.5 | 88.2 | 0.4 (0.2–0.8) | 0.5 (0.3–1.1) |
| Eat more olive oil | 26.6 | 34 | 0.7 (0.5–1.1) | 0.7 (0.5–1.2) | 29.4 | 37.9 | 0.7 (0.4–1.1) | 0.8 (0.5–1.5) |
| Use sunblock or sunscreen | 26 | 50.8 | 0.3 (0.2–0.5) | 0.3 (0.2–0.5) | 32 | 44.4 | 0.6 (0.4–0.9) | 0.8 (0.4–1.3) |
| Cut down on alcohol consumption | 11.9 | 25.3 | 0.4 (0.2–0.7) | 0.4 (0.2–0.7) | 20.9 | 22.9 | 0.9 (0.5–1.6) | 0.7 (0.4–1.3) |
| Cut down on caffeine | 33 | 34.3 | 0.9 (0.6–1.4) | 0.9 (0.6–1.5) | 29.6 | 34 | 0.8 (0.5–1.3) | 0.8 (0.4–1.3) |
| Take vitamin supplements | 47.6 | 55.6 | 0.7 (0.5–1.1) | 0.8 (0.5–1.3) | 48.7 | 60.8 | 0.6 (0.4–0.96) | 0.8 (0.4–1.3) |
| Meditate or use other technique to reduce stress | 22.2 | 38.4 | 0.5 (0.3–0.7) | 0.5 (0.3–0.9) | 22.2 | 35.3 | 0.5 (0.3–0.9) | 0.6 (0.3–1.03) |
| Cut down on red meat | 56.2 | 72.7 | 0.5 (0.3–0.7) | 0.5 (0.3–0.8) | 62.1 | 71.9 | 0.6 (0.4–1.03) | 0.6 (0.3–0.97) |
| Take calcium supplement | 32.6 | 42.9 | 0.6 (0.4–0.98) | 0.9 (0.5–1.4) | 30.5 | 51.3 | 0.4 (0.3–0.7) | 0.5 (0.3–0.9) |
| Take fish oil supplement | 4.3 | 8.6 | 0.5 (0.2–1.2) | 0.4 (0.2–1.1) | 5.2 | 6.5 | 0.8 (0.3–2.1) | 0.7 (0.2–2.1) |
| Any other thing to try to stay healthy | 23.2 | 24.2 | 0.9 (0.6–1.5) | 1.1 (0.7–1.8) | 21.6 | 23.5 | 0.9 (0.5–1.5) | 0.9 (0.5–1.7) |

[1]age, hypertension, diabetes, body mass index (quartiles), smoking, race and ethnicity, level of education and history of coronary heart disease.
[2]age, hypertension, diabetes, body mass index (quartiles), smoking, race and ethnicity, level of education and history of stroke/TIA.

TABLE 4
*Women with hysterectomy: percent of cases and controls who reported the behavior and adjusted odds ratios for myocardial infarction and stroke in women who reported the behavior*

| Health behaviors | Myocardial infarction | | | | Stroke | | | |
|---|---|---|---|---|---|---|---|---|
| | Cases (N = 125) | Controls (N = 122) | Age adjusted OR (95% CI) | Adjusted for traditional confounders[1] OR (95% CI) | Cases (N = 90) | Controls (N = 85) | Age adjusted (95% CI) | Adjusted for traditional confounders[2] OR (95% CI) |
| Exercise | 54.5 | 69.7 | 0.5 (0.3–0.9) | 0.6 (0.3–1.06) | 62.1 | 78.8 | 0.4 (0.2–0.9) | 0.5 (0.2–1.2) |
| Try to eat more foods containing fiber | 81.5 | 85.3 | 0.8 (0.4–1.5) | 0.8 (0.4–1.7) | 80.7 | 89.4 | 0.5 (0.2–1.2) | 0.5 (0.2–1.6) |
| Try to eat foods low in fat | 82.3 | 86.1 | 0.8 (0.4–1.5) | 0.9 (0.4–1.8) | 83 | 89.4 | 0.6 (0.2–1.5) | 0.5 (0.2–1.5) |
| Eat more olive oil | 29 | 32.5 | 0.8 (0.5–1.5) | 1.0 (0.5–1.9) | 30 | 43.5 | 0.6 (0.3–1.04) | 0.8 (0.4–1.8) |
| Use sunblock or sunscreen | 37.1 | 48.4 | 0.6 (0.4–1.04) | 0.8 (0.4–1.4) | 27.3 | 57.7 | 0.3 (0.1–0.5) | 0.4 (0.2–0.9) |
| Cut down on alcohol consumption | 22.6 | 30.3 | 0.7 (0.4–1.2) | 0.6 (0.3–1.2) | 18.2 | 25.9 | 0.6 (0.3–1.3) | 0.7 (0.3–1.7) |
| Cut down on caffeine | 32 | 45.1 | 0.6 (0.3–0.96) | 0.6 (0.3–1.00) | 31.8 | 37.7 | 0.8 (0.4–1.5) | 0.9 (0.4–2.0) |
| Take vitamin supplements | 50.8 | 61.5 | 0.6 (0.4–1.08) | 0.6 (0.4–1.08) | 54.6 | 64.7 | 0.7 (0.4–1.2) | 0.7 (0.3–1.5) |
| Meditate or use other technique to reduce stress | 28.2 | 41.8 | 0.5 (0.3–0.9) | 0.5 (0.3–0.9) | 29.9 | 47.1 | 0.5 (0.3–0.9) | 0.5 (0.2–1.04) |
| Cut down on red meat | 63.7 | 74.6 | 0.6 (0.3–1.04) | 0.5 (0.3–0.97) | 67.1 | 78.8 | 0.6 (0.3–1.1) | 0.5 (0.2–1.1) |
| Take calcium supplement | 39.5 | 51.6 | 0.6 (0.4–1.02) | 0.8 (0.5–1.4) | 44.3 | 47.1 | 0.9 (0.5–1.7) | 1.1 (0.5–2.2) |
| Take fish oil supplement | 4.8 | 7.4 | 0.6 (0.2–1.9) | 0.9 (0.3–3.1) | 6.8 | 7.1 | 1.0 (0.3–3.2) | 0.9 (0.2–3.7) |
| Any other thing to try to stay healthy | 24.2 | 23.1 | 1.1 (0.6–1.9) | 1.2 (0.6–2.4) | 27.6 | 29.4 | 0.9 (0.5–1.8) | 1.1 (0.5–2.6) |

[1] age, hypertension, diabetes, body mass index (quartiles), smoking, race and ethnicity, level of education and history of coronary heart disease.
[2] age, hypertension, diabetes, body mass index (quartiles), smoking, race and ethnicity, level of education and history of stroke/TIA.

TABLE 5

Odds ratios for myocardial infarction and ischemic stroke in current users of estrogen alone or estrogen/progestin after various adjustments and comparison with women's health initiative clinical trial results

| Adjustment | Women without hysterectomy / estrogen plus progestin | | Women with hysterectomy / estrogen only | |
|---|---|---|---|---|
| | Myocardial infarction odds ratio (95% C.I.) | Ischemic stroke odds ratio (95% C.I.) | Myocardial infarction odds ratio (95% C.I.) | Ischemic stroke odds ratio (95% C.I.) |
| **Current analysis** | | | | |
| Age only | 0.62 (0.38–1.007) | 0.71 (0.41–1.24) | 0.71 (0.41–1.25) | 0.72 (0.37–1.42) |
| Age and traditional confounders[1] | 0.91 (0.52–1.58) | 0.85 (0.43–1.66) | 0.96 (0.50–1.82) | 1.17 (0.48–2.90) |
| Age, traditional confounders[1] + health behavior variables selected using stepwise regression | 1.13 (0.62–2.06)[2] | 1.00 (0.50–2.01)[3] | 1.04 (0.52–2.00)[4] | 1.01 (0.40–2.54)[5] |
| **Women's Health Initiative clinical trial** | 1.32 (1.02–1.72)[6] | 1.41 (1.07–1.85)[7] | 0.89 (0.70–1.12)[6] | 1.39 (1.10–1.77)[7] |

[1]Age, hypertension, diabetes, body mass index, smoking, race and ethnicity, level of education, history of coronary heart disease for myocardial infarction and history of stroke for stroke.
[2]Try to eat more foods low in fat; use sunblock or sunscreen; cut down on alcohol consumption; cut down on caffeine; meditate or use other technique to reduce stress; cut down red meat; take fish oil supplement.
[3]Exercise; cut down on red meat; take calcium supplement.
[4]Cut down on caffeine; take vitamin supplements; meditate or use other technique to reduce stress; cut down red meat.
[5]Use sunblock or sunscreen; meditate or use other technique to reduce stress.
[6]Non-fatal MI; nominal confidence interval.
[7]Fatal plus non-fatal; nominal confidence interval.

less than 1.0. Of particular note is the persistence of the association of sunscreen use with lower risk of MI and stroke even after adjustment for confounders. It is highly unlikely that sunscreen use prevents MI or stroke. We chose to query women in the study about this behavior precisely because there was no immediately plausible direct causal pathway between sunscreen use and these cardiovascular endpoints. We hypothesized a priori that sunscreen use and the other health behaviors would be markers of a "healthy lifestyle" and that adjustment for a measure of healthy lifestyle would improve inferences about the effect of hormone therapy on MI and stroke.

Our findings with regard to sunscreen use shows that a strategy in which non-causal variables are systematically measured in appropriate models might improve inferences deriving from observational research. Education, social status, and income are examples of variables that epidemiologists frequently consider as confounders even though their relationship with disease is seldom causal. Rather, the variables "capture" the causal associations of factors associated with the exposure that get "mixed" with the true effect of the exposure.

Our analysis does not include some variables that are measurable (family history of heart disease, untreated hypertension) and whose inclusion might further change the estimates. We did not include interaction terms in the models and this might also have brought the data closer to WHI. Where to stop when adjusting is generally left to the researcher's judgment. When the "truth" is known, it may be possible to find the "truth." The problems arise when the researcher seeks the truth through modeling.

In the current analysis, adjustment for the behaviors with the strongest association with the given vascular endpoint moved the ORs estimates for current E and E/P closer to values observed in the WHI. Of course, we know that the adjustment moved the ORs in the "right" direction only because of the clinical trial. Even with the adjustment for healthy behavior/lifestyle in addition to adjustment for known confounders, we were unable to reproduce completely the clinical trial results.

**Acknowledgments.** Thanks to Steve Sidney who was a co-investigator in the main study.